\documentclass[]{aastex631}

\begin{document}

\title{Case A or Case B? The effective recombination coefficient in\\ gas clouds of arbitrary optical thickness}

\author{Olof Nebrin}
\affiliation{Department of Astronomy \& Oskar Klein Centre, \\
AlbaNova, Stockholm University, \\
SE-106 91 Stockholm, Sweden}



\begin{abstract}
In calculations of the ionization state, one is often forced to choose between the Case A recombination coefficient $\alpha_{\rm A}$ (sum over recombinations to all hydrogen states) or the Case B recombination coefficient $\alpha_{\rm B}$ (sum over all hydrogen states except the ground state). If the cloud is optically thick to ionizing photons, $\alpha_{\rm B}$ is usually adopted on the basis of the "on-the-spot" approximation, wherein recombinations to the ground state are ignored because they produce ionizing photons absorbed nearby. In the opposite case of an optically thin cloud, one would expect the Case A recombination coefficient to better describe the effective recombination rate in the cloud. In this paper, I derive an analytical expression for the effective recombination coefficient in a gas cloud of arbitrary optical thickness which transitions from $\alpha_{\rm A}$ to $\alpha_{\rm B}$ as the optical thickness increases. The results can be readily implemented in numerical simulations and semi-analytical calculations. 

\end{abstract}

%%\keywords{Classical Novae (251) --- Ultraviolet astronomy(1736) --- History of astronomy(1868) --- Interdisciplinary astronomy(804)}


\section{Introduction} 
\label{sec:intro}

If the diffuse ionizing radiation is not solved for in detail, one is forced to choose between the Case A and Case B recombination coefficients to model the recombination rate of ionized gas. The wrong choice can lead to significant errors. \cite{Milan2014} have shown that adopting either the Case A or Case B recombination coefficient, without modelling the diffuse radiation, can lead to significant inaccuracies in the modelling of H II regions around stars, as well as the ionization state of the 'shadows' behind illuminated optically thick absorbers. Significant errors can also appear in cosmological settings. For example, wrongly adopting the Case A instead of the Case B recombination coefficient -- as done by some authors -- in the study of chemistry in pristine atomic-cooling halos at Cosmic Dawn can lead to an error of $80-90\%$ in the critical Lyman-Werner intensity needed to form direct-collapse black holes \citep{Glover2015}, which in turn could lead to large errors in their predicted abundance. Similarly, choosing either $\alpha_{\rm A}$ or $\alpha_{\rm B}$ can lead to shifts of $\Delta z \sim 0.5$ in the predicted end of reionization in models, all else being equal \citep{Kaurov2014}. It would be useful to have a general expression for the effective recombination coefficient $\alpha_{\rm eff}$ that interpolates between Case A and Case B as the cloud optical thickness increase. \cite{Davidson1977} proposed a crude "modified on-the-spot" method for $\alpha_{\rm eff}$ which does this \citep[see also, e.g.,][]{Netzer1984, Netzer1990}, but it was for slab geometry and no derivation was given for the result. Below I derive the effective recombination coefficient for a spherical gas cloud of uniform density and ionization state and arbitrary optical thickness. This is done by solving for the diffuse ionizing radiation using the Eddington approximation. The final result (Eqs. \ref{alphaeff} and \ref{alpha_eff_av}) can be applied in e.g. semi-analytical models of gas chemistry in collapsing regions, or on a cell-to-cell basis in numerical simulations of reionization.

\section{Solving the radiative transfer equation} \label{sec:radtransf}
Consider a stationary spherical gas cloud of radius $R$, uniform hydrogen density $n_{\rm H} = n_{\rm H^0} + n_{\rm H^+}$ and ionization fraction $x \equiv n_{\rm H^+}/n_{\rm H}$. To get the effective recombination coefficient we need to know the ionization rate as a result of recombination coefficients to the ground state. This in turn requires us to solve for the diffuse ionizing radiation within the cloud. Thus, we start with the radiative transfer equation in spherical coordinates \citep[e.g.,][]{Chandrasekhar1960}:
\begin{equation}
    \mu \frac{\partial I_{\nu}}{\partial r} + \frac{1 - \mu^2}{r} \frac{\partial I_{\nu}}{\partial \mu} = j_{\rm fb, \nu} - n_{\rm H^0} \sigma_{\nu} I_{\nu} \hspace{1 pt} .
    \label{radtransf}
\end{equation}
Here $\sigma_{\nu}$ is the hydrogen photoionization cross-section, and $j_{\rm fb, \nu}$ the emissivity due to recombinations to the ground state:
\footnote{The ground state recombination line has a spectrum $\propto e^{(E_{\rm LyC} - h\nu)/kT}$, so approximating the recombination line as a Dirac delta function is a good approximation for temperatures $T \lesssim 0.1 E_{\rm LyC}/k \simeq  1.6 \times 10^4 \hspace{1 pt} \rm K$.}
\begin{equation}
    j_{\rm fb, \nu} \simeq \frac{\alpha_{1} n_{\rm H^+} n_{\rm e}}{4 \pi} \hspace{1 pt} E_{\rm LyC} \hspace{1 pt} \delta_{\rm D}(\nu_{\rm LyC}) \hspace{1 pt} ,
    \label{emissivity}
\end{equation}
where $E_{\rm LyC} = h\nu_{\rm LyC} = 13.6 ~ \rm eV$ is the ionization threshold, and $\alpha_1 = \alpha_{\rm A} - \alpha_{\rm B}$ is the recombination rate to the ground state. We can solve Eq. (\ref{radtransf}) using the Eddington approximation, wherein the intensity is assumed to be nearly isotropic. This approximation is expected to be good in this case because the emissivity in Eq. (\ref{emissivity}) is isotropic. The zeroth, first, and second moments of the intensity are:
\begin{equation}
    J_{\nu} = \frac{1}{2}\int_{-1}^{1} \textrm{d}\mu \hspace{1 pt} I_{\nu} \hspace{1 pt}, ~~ H_{\nu} = \frac{1}{2}\int_{-1}^{1} \textrm{d}\mu \hspace{1 pt} \mu I_{\nu} \hspace{1 pt}, ~~ K_{\nu} = \frac{1}{2}\int_{-1}^{1} \textrm{d}\mu \hspace{1 pt} \mu^2 I_{\nu} \hspace{1 pt}.
\end{equation}
In the Eddington approximation one make the linear approximation $I_{\nu}(r,\mu) \simeq a_{\nu}(r) + b_{\nu}(r) \mu$, which leads to the closure relation $K_{\nu} = J_{\nu}/3$ \citep[e.g.,][]{Rybicki1986}. Taking the zeroth and first moments of Eq. (\ref{radtransf}) and employing the closure relation yields:
\begin{equation}
    \frac{\textrm{d} H_{\nu}}{\textrm{d}r} + \frac{2 H_{\nu}}{r} = j_{\rm fb, \nu} -  n_{\rm H^0} \sigma_{\nu} J_{\nu} \hspace{1 pt}, ~~ \frac{1}{3} \frac{\textrm{d} J_{\nu}}{\textrm{d}r} = - n_{\rm H^0} \sigma_{\nu} H_{\nu}  \hspace{1 pt}.
    \label{twodiffeqs}
\end{equation}
These equations can be combined into a single one for $J_{\nu}$:
\begin{equation}
    \frac{\textrm{d}^2 J_{\nu}}{\textrm{d} \tau_\nu^2} + \frac{2}{\tau_\nu}  \frac{\textrm{d} J_{\nu}}{\textrm{d} \tau_\nu} - 3 J_\nu = - \frac{3 j_{\rm fb,\nu}}{n_{\rm H^0} \sigma_\nu} \hspace{1 pt} ,
\end{equation}
where I have introduced the optical depth from the center, $\tau_\nu(r) = \int_0^r \textrm{d}r' \hspace{1 pt} n_{\rm H^0} \sigma_\nu$ (so that it is zero at the center). The above differential equation can be solved using the simplifying assumptions of uniform density and ionization fraction. The general solution is:
\begin{equation}
    J_\nu(\tau_\nu) =  \frac{j_{\rm fb,\nu}}{n_{\rm H^0} \sigma_\nu} + \frac{1}{\tau_\nu} \left(A e^{-\sqrt{3}\tau_\nu} + B e^{\sqrt{3}\tau_\nu} \right ) \hspace{1 pt} .
\end{equation}
The constants $A$ and $B$ can be determined from the boundary conditions:
\begin{enumerate}
    \item Zero net flux $F_{\nu} = 4 \pi H_\nu$ at the center ($\tau_\nu = 0$).
    \item No incoming diffuse intensity $I_\nu^-$ at the cloud edge (at optical depth $\tau_{\rm cl, \nu} \equiv n_{\rm H^0} \sigma_\nu R$).
\end{enumerate}
The first boundary condition implies that $\textrm{d}J_\nu/\textrm{d}\tau_\nu = 0$ at $\tau_\nu = 0$ (see Eq. \ref{twodiffeqs}). The second boundary condition can be treated with the two-stream approximation \citep[e.g.,][]{Rybicki1986}, where the intensity is assumed to travel at two angles $\mu = +1/\sqrt{3}$ (outgoing) and $\mu = -1/\sqrt{3}$ (incoming). With our convention for the optical depth increasing from the center, this yields $I_\nu^- = J_\nu + (1/\sqrt{3})(\textrm{d}J_\nu/\textrm{d}\tau_\nu)$. This should be zero at $\tau_{\rm cl, \nu}$ to satisfy the second boundary condition. After some algebra and using Eq. (\ref{emissivity}) one ends up with the solution:
\begin{equation}
    J_{\nu}(\tau_\nu) = \frac{\alpha_1 n_{\rm H^+} n_{\rm e} E_{\rm LyC}}{4 \pi n_{\rm H^0} \sigma_\nu} \delta_{\rm D} (\nu_{\rm LyC}) \Biggl\{ 1 + \frac{1}{\tau_\nu} \left( e^{-\sqrt{3}\tau_\nu} - e^{\sqrt{3}\tau_\nu} \right ) \left[ \frac{2}{\tau_{\rm cl,\nu}} e^{\sqrt{3}\tau_{\rm cl,\nu}} + \frac{1}{\sqrt{3}\tau_{\rm cl,\nu}^2} \left( e^{-\sqrt{3}\tau_{\rm cl,\nu}} - e^{\sqrt{3}\tau_{\rm cl,\nu}} \right)  \right]^{-1}  \Biggl\} \hspace{1 pt} .
    \label{Jsolution}
\end{equation}
This solution will determine the photoionization rate from the diffuse ionizing radiation in the cloud.

\section{The effective recombination coefficient} \label{sec:effrecomb}

\begin{figure}
\plotone{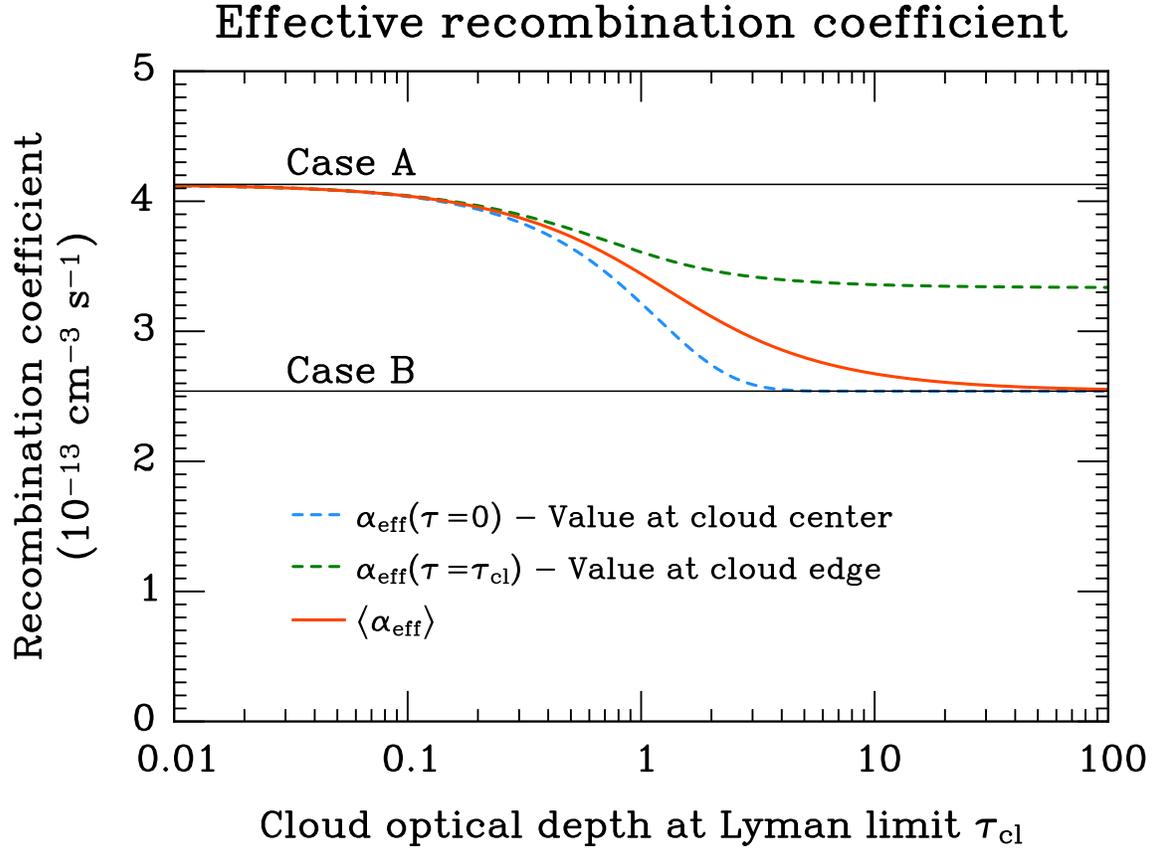}
\caption{The predicted effective recombination coefficient at the cloud center ($ \alpha_{\rm eff}(\tau = 0)$), the cloud edge ($ \alpha_{\rm eff}(\tau = \tau_{\rm cl})$), and the cloud-averaged value ($ \langle \alpha_{\rm eff} \rangle$). The Case A and Case B recombination coefficients were computed for $T = 10^4 \hspace{1 pt} \rm K$ \citep{Draine2011}.}
\label{fig:result}
\end{figure}

With the solution for the diffuse mean ionizing intensity $J_\nu$ we can now determine the effective recombination coefficient. We can define it from the time evolution of the ionized hydrogen number density:
\begin{equation}
    \frac{\partial n_{\rm H^+}}{\partial t} = \Gamma_{\star} n_{\rm H^0} +  \Gamma_{\rm diff} n_{\rm H^0} -\alpha_{\rm A} n_{\rm H^+} n_{\rm e} \equiv \Gamma_{\star} n_{\rm H^0} - \alpha_{\rm eff} n_{\rm H^+} n_{\rm e} \hspace{1 pt} ,
\end{equation}
where $\Gamma_{\rm diff} = \int_{\nu_{\rm LyC}}^{\infty} \textrm{d}\nu \hspace{1 pt} 4 \pi J_\nu \sigma_\nu /h\nu$ is the photoionization rate due to the diffuse ionizing radiation, and $\Gamma_{\star}$ the photoionization rate from any luminous sources (e.g. nearby stars). The above definition yields an effective recombination coefficient of:
\begin{equation}
    \alpha_{\rm eff} = \alpha_{\rm A} - \frac{\Gamma_{\rm diff} n_{\rm H^0}}{n_{\rm H^+} n_{\rm e}} \hspace{1 pt} .
\end{equation}
Using Eq. (\ref{Jsolution}) we find a diffuse photoionization rate of
\begin{equation}
    \Gamma_{\rm diff}(\tau) =  \frac{\alpha_1 n_{\rm H^+} n_{\rm e}}{n_{\rm H^0} }  \Biggl\{ 1 + \frac{1}{\tau} \left( e^{-\sqrt{3}\tau} - e^{\sqrt{3}\tau} \right ) \left[ \frac{2}{\tau_{\rm cl}} e^{\sqrt{3}\tau_{\rm cl}} + \frac{1}{\sqrt{3}\tau_{\rm cl}^2} \left( e^{-\sqrt{3}\tau_{\rm cl}} - e^{\sqrt{3}\tau_{\rm cl}} \right)  \right]^{-1}  \Biggl\} \hspace{1 pt} ,
\end{equation}
where $\tau$ and $\tau_{\rm cl}$ are $\tau_{\nu}$ and  $\tau_{\rm cl, \nu}$ evaluated at the Lyman limit ($\nu_{\rm LyC}$), respectively. Thus, the effective recombination coefficient at position $\tau$ becomes (using $\alpha_1 = \alpha_{\rm A} - \alpha_{\rm B}$):
\begin{equation}
    \alpha_{\rm eff}(\tau) = \alpha_{\rm A} -  (\alpha_{\rm A} - \alpha_{\rm B})  \Biggl\{ 1 + \frac{1}{\tau} \left( e^{-\sqrt{3}\tau} - e^{\sqrt{3}\tau} \right ) \left[ \frac{2}{\tau_{\rm cl}} e^{\sqrt{3}\tau_{\rm cl}} + \frac{1}{\sqrt{3}\tau_{\rm cl}^2} \left( e^{-\sqrt{3}\tau_{\rm cl}} - e^{\sqrt{3}\tau_{\rm cl}} \right)  \right]^{-1}  \Biggl\} \hspace{1 pt} \label{alphaeff}.
\end{equation}
This is dependent on the position within the cloud. The cloud-averaged recombination coefficient that determines the recombination rate of the cloud as a whole is given by:
\begin{equation}
    \langle \alpha_{\rm eff} \rangle = \frac{\int_0^R \textrm{d}r \hspace{1 pt} \alpha_{\rm eff}(r) n_{\rm H^+} n_{\rm e} r^2}{\int_0^R \textrm{d}r \hspace{1 pt}  n_{\rm H^+} n_{\rm e} r^2} = \frac{3}{\tau_{\rm cl}} \int_0^{\tau_{\rm cl}} \textrm{d}\tau \hspace{1 pt} \alpha_{\rm eff}(\tau) \tau^2 \hspace{1 pt} ,
\end{equation}
where I have used the assumption of homogeneity to get the final integral over optical depth. Using Eq. (\ref{alphaeff}) the average effective recombination coefficient becomes:
\begin{equation}
    \langle \alpha_{\rm eff} \rangle = \alpha_{\rm A} -  (\alpha_{\rm A} - \alpha_{\rm B})f(\tau_{\rm cl}) \hspace{1 pt} ,
    \label{alpha_eff_av}
\end{equation}
where the function $f(\tau_{\rm cl})$ is given by
\begin{equation}
    f(\tau_{\rm cl}) = 1 - \frac{1 + \sqrt{3}\tau_{\rm cl} - e^{2\sqrt{3}\tau_{\rm cl}}\left(1 - \sqrt{3}\tau_{\rm cl} \right) }{2 \tau_{\rm cl}^2 e^{2\sqrt{3}\tau_{\rm cl}} + \frac{\tau_{\rm cl}}{\sqrt{3}} \left( 1 -  e^{2\sqrt{3}\tau_{\rm cl}} \right) } \hspace{1 pt} .
\end{equation}
The effective recombination rates $\langle \alpha_{\rm eff} \rangle$, $\alpha_{\rm eff}(\tau = 0)$, and $\alpha_{\rm eff}(\tau = \tau_{\rm cl})$ are plotted in Fig. \ref{fig:result}. For gas clouds that are optically thin to ionizing photons ($\tau_{\rm cl} \ll 1$) we have $f(\tau_{\rm cl}) \ll 1$ and hence $ \langle \alpha_{\rm eff} \rangle \simeq \alpha_{\rm A}$. For optically thick clouds ($\tau_{\rm cl} \gg 1$) we instead have $f(\tau_{\rm cl}) \simeq 1$ and so $\langle \alpha_{\rm eff} \rangle \simeq \alpha_{\rm B}$. The cloud-averaged effective recombination coefficient, therefore, interpolate between Case A and Case B as expected. The same is true for the effective recombination coefficient at the center of the cloud, $\alpha_{\rm eff}(\tau = 0)$. The optically thick limit for $\alpha_{\rm eff}(\tau = \tau_{\rm cl})$ falls right in the middle between Case A and Case B because half of the diffuse ionizing photons produced at the cloud edge can freely escape the cloud \citep[as noted by][]{Davidson1977, Netzer1984}. 

%% Also note that the akcnowlodgment environment does not support long amounts of text. If you have a lot of people and institutions to acknowledge, do not use this command. Instead, create a new \section{Acknowledgments}.
\begin{acknowledgments}
I want to thank Garrelt Mellema, Sambit K. Giri, and the Stockholm Reionization Group at large for interesting and helpful discussions.
\end{acknowledgments}






\bibliography{bibliography}{}
\bibliographystyle{aasjournal}



\end{document}